\documentclass[final,5p,times]{elsarticle}
\usepackage[utf8]{inputenc}
\usepackage{siunitx}
\usepackage{hyperref}
\usepackage{graphicx}
\usepackage{chngcntr}
\usepackage{amsmath}
\usepackage{color}
\usepackage{soul}
\usepackage[caption=false]{subfig}
\title{Feed-forward neural network unfolding}
\date{}

\begin{document}
\author[1]{Ming-Liang Wong}
\author[2]{Andrew Edmonds}
\author[3]{Chen Wu}

\address[1]{M. Smoluchowski Institute of Physics, Jagiellonian University, Krak\'{o}w, Poland}
\address[2]{Boston University, Boston, Massachusetts, USA}
\address[3]{Osaka University, Osaka, Japan}

\begin{abstract}
A feed-forward neural network is demonstrated to efficiently unfold the energy distribution of protons and alpha particles passing through passive material. This model-independent approach works with unbinned data and does not require regularization. The training dataset was produced with the same Monte Carlo simulation framework used by the AlCap experiment. The common problem of designing a network is also addressed by performing a hyperparameter space scan to find the best network geometry possible within reasonable computation time. Finally, a comparison with other unfolding methods such as the iterative d'Agostini Bayesian unfolding, and Singular Value Decomposition (SVD) are shown.
\end{abstract}
\maketitle

\section{Introduction}
When measuring particle energies, a detector will distribute those energies over many channels determined by its acceptance and resolution functions. This detector energy smearing must be modelled by a thorough study of detector response as we are interested only in the true energy distribution. The direct method would be to compare the convolution of the detector response and the true energy distribution generated with Monte Carlo simulation with detector measured energies. However, this might be impossible since we do not have sufficient knowledge of the detector response and the energy spectra we are measuring. The inverse method would then be to remove such convolutions from data by employing an unfolding technique. This difficult process of unfolding is typically necessary to combine results from different experiments and to tune Monte Carlo simulations since it remove effects from detectors.

In recent years, machine learning algorithms such as neural networks have found applications in many areas of nuclear and particle physics. Tools have been created for detector response unfolding\cite{AVDIC2006742}, neutron spectrum unfolding\cite{HOSSEINI201675}, pulse height and pile-up analysis\cite{REGADIO2021165403}, and energy spectrum unfolding\cite{gagunashvili2011machine}. In this paper, a feed-forward neural network (NN) is trained to recover the initial or truth distribution, $E_i$ from detector measured energies, $E_f$. The NN is not without its challenges, one of them being that having hidden layers with a non-convex loss function may lead to different validation accuracy depending on their weight initialisation. Insufficient independently labelled and good quality data for use in training would also produce questionable results. In addition, it also requires tuning of hyper-parameters such as the number of nodes or layers which constitute the geometry of the network and training iterations (epochs) to obtain the best results. We will address this by performing a brute force scan of the hyper-parameter space which should be sufficient to provide an example to show that the performance of NN unfolding is on par if not better than current techniques. Quality checks on the training data were done to reduce uncertainties due to mis-modeling and logical errors.

\section{Monte Carlo simulation}
\begin{figure}[ht]
    \centering
    \includegraphics[width=.9\columnwidth]{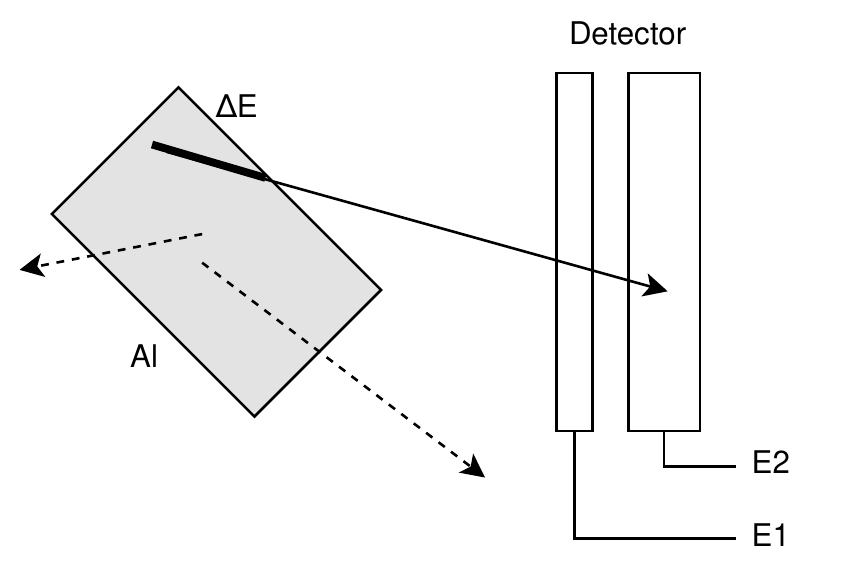}
    \caption{In addition to performing particle identification, the dE/dx Si-detector measures the energy deposited by protons, alpha particles or any other charged particle. However, the particles loses some unknown energy, $\Delta E$ while traversing through some aluminium. The figure above is not to scale.}
    \label{fig:p-E-loss}
\end{figure}

One of the primary aims of AlCap\cite{alcapcollaboration2021measurement} was to measure the charged particle emission rate after nuclear muon capture on aluminium. The experiment stopped muons in an aluminium target which then underwent the capture process and produced a myriad of particles such as protons and alpha particles. These particles will lose some energy $\Delta E \propto 1/E$ when passing through the thin aluminium target and a small fraction of them will end up in the dE/dx silicon detector placed \SI{122}{mm} away from the aluminium as shown in Fig.\ref{fig:p-E-loss}.

The training data for the NN was obtained from the Monte Carlo simulation framework developed for the AlCap experiment. In this framework, we generated response matrices which were used to unfold the raw spectra with the d'Agostini iterative method. However, this method required regularisation which is the number of iterations and there was no convincing way to determine this number. We used four iterations but assigned a systematic uncertainty to it by comparing the unfolded result with 20 iterations. The response matrices were also being cut off abruptly because of detection limits, which created large un-physical fluctuations near the low and high energy regions due to low statistics and low probability of escaping the target.

For this paper, we generated a total of $2\times 10^7$ particles with their initial positions Gaussian-distributed from the middle of a thin piece of aluminium. They were generated with a uniform initial energy distribution, $E_i$ between 3 and \SI{20}{MeV} for protons and between 0 and \SI{60}{MeV} for alphas and emitted isotropically.  A uniform distribution is selected to minimize the bias to any type of distribution the NN might encounter during unfolding.
The detector itself consists of one thin and one thick silicon sub-detector to measure the total energy deposited by incident charged particles. The thin sub-detector records the energy deposited as $E_1$ and the thick sub-detector as $E_2$ and their sum is the measured energy of a particles travelling out from the aluminium. Passive layers of \SI{0.5}{\mu m} passive layers are also present on both sub-detectors as well as minimum energy thresholds of \SI{0.2}{MeV} to mimic real detector conditions. We consider only particles that record at least one hit in both the thin and thick sub-detector and deposits all its remaining kinetic energy in the thick sub-detector. We can write this effect as
\begin{equation}
    E_f = M_{fi} E_i
\end{equation}
where the matrix $M_{fi}$ contains both the energy loss of the proton due to the aluminium target and the dE/dx Si-detector effects. $E_f$ would be the final measured energies as the sum $E_1 + E_2$.

\begin{figure}[ht]
    \subfloat[Protons ]{%
        \includegraphics[width=.9\columnwidth]{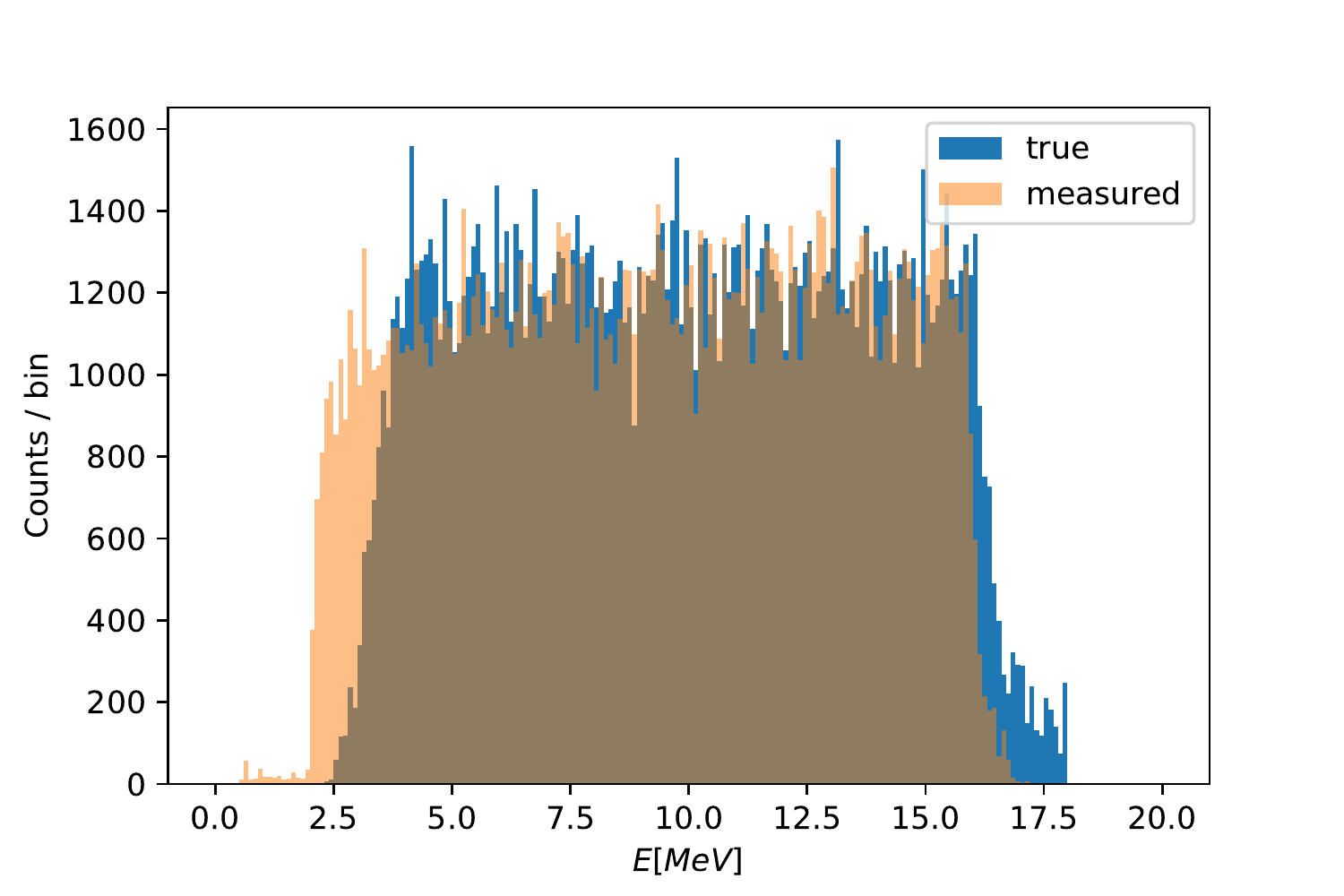}%
    }
    
    \subfloat[Alpha particles]{%
        \includegraphics[width=.9\columnwidth]{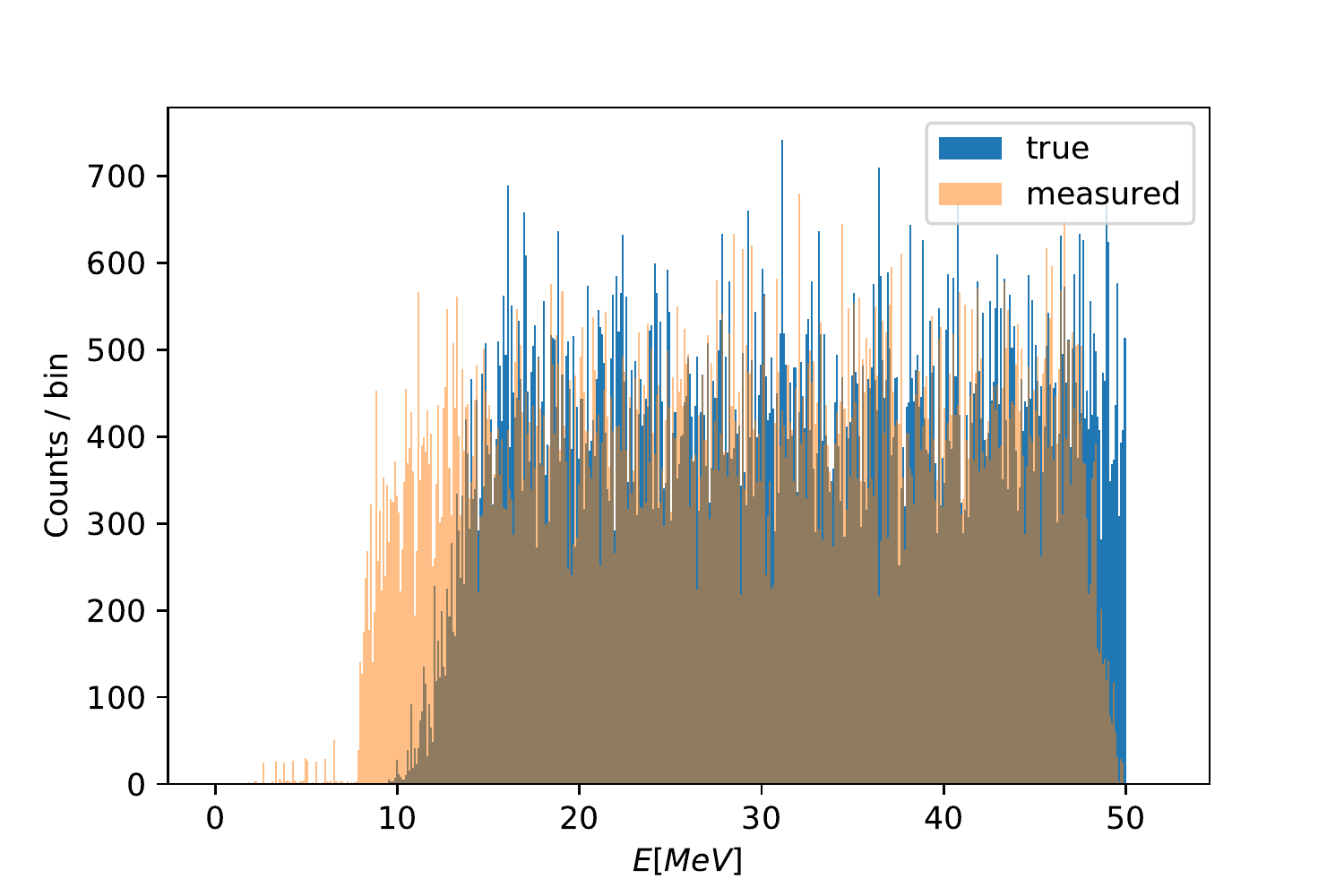}%
    }
    \caption{The proton and alpha energy distribution as measured by the silicon detector clearly shows a shift to lower energies as compared to the true values due to energy loss in the aluminium target. Bin width is \SI{0.1}{MeV}. }
    \label{fig:proton_energy-distribution-sir}
\end{figure}

 Fig.\ref{fig:proton_energy-distribution-sir} shows the proton energy distribution as measured by the detector. The energy cutoff at about \SI{17}{MeV} is due to rejection of events with protons not stopping in the thick subdetector.

\section{Setup and training}
\begin{figure}[ht]
    \centering
    \subfloat[Protons]{\includegraphics[width=.9\columnwidth]{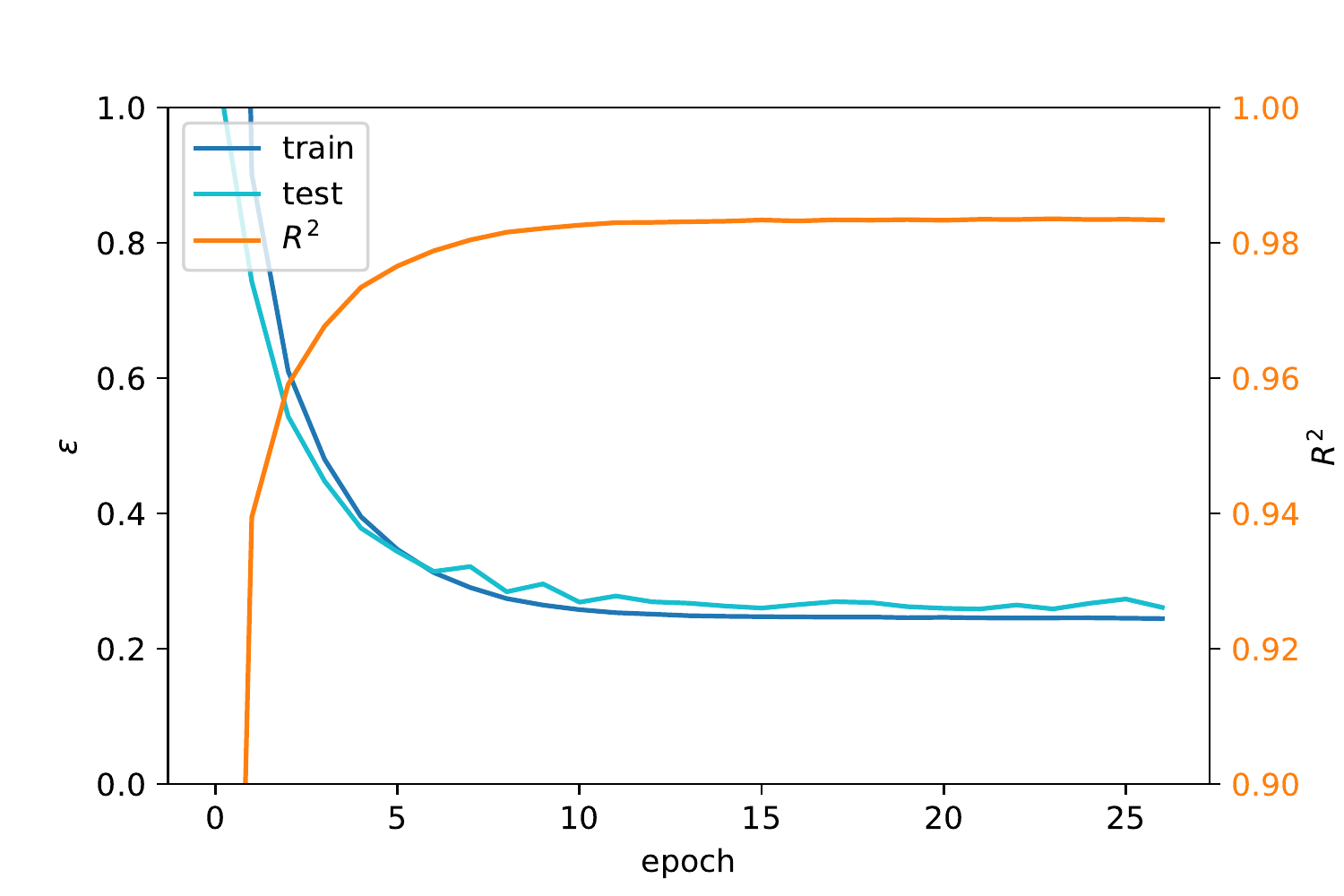} }
    
    \subfloat[Alpha particles]{\includegraphics[width=.9\columnwidth]{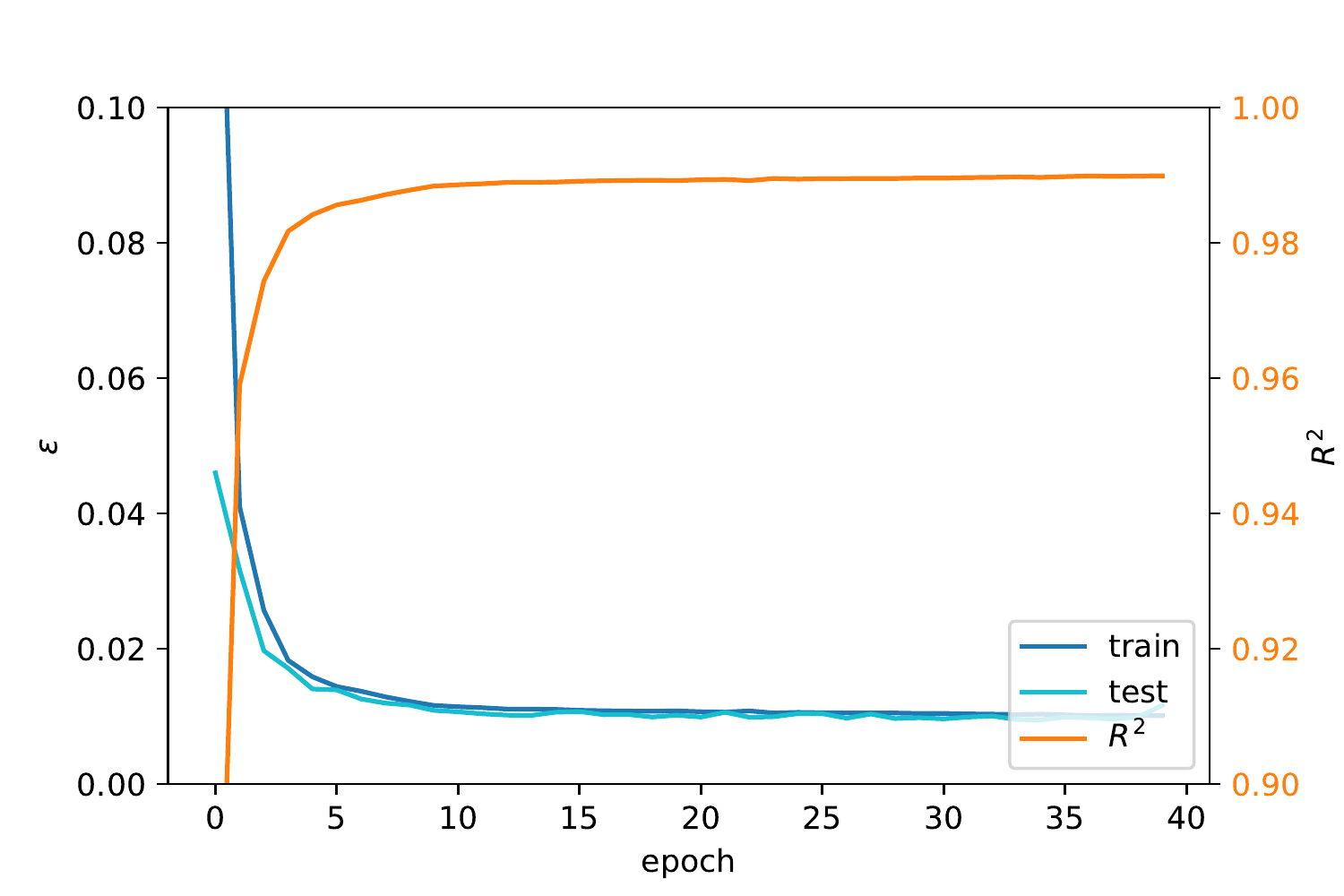} }
    \caption{The $R^2$ loss value, $\epsilon$ decreasing quickly which typically shows high learning rate and the $R^2$ trend also shows high accuracy of up to  $>0.98$ with no visible overfitting.}
    \label{fig:loss-vs-epoch}
\end{figure}

A simple NN with two input nodes with values from each subdetector, $E_1$ and $E_2$ is connected to one hidden layer with 24 nodes for protons and alphas. This geometry was found to be optimal within the hyperparameter search space as described in Sec~\ref{sec:hyperparameter_tuning}. As unfolding a physical quantity is effectively a regression problem, the output layer will only consist of one node which provides the energy correction factor. In the hidden layers, the Rectified Linear Unit, (ReLU) activation function, $\sigma = \texttt{max}(0, z) $ is used. The Keras API\cite{chollet2015keras} is used to construct the NN. The network uses an extension to the stochastic gradient descent algorithm, Adam\cite{kingma2017adam} learning optimizer which sets the learning rate to 0.001, $\beta_1 = 0.9$, $\beta_2 = 0.9$, $\epsilon = 10^{-7}$, with no AMSGrad\cite{reddi2019convergence} variant by default. Eq.\ref{eqn:nn-matrix} describes the result of the NN which are weights, $\mathcal{W}$ and biases, $\mathcal{B}$ that are used for predicting the unfolded energies, $\hat{y}$ given the inputs $\hat{x} = [E_1, E_2]$ and the choice of an activation function $\sigma_{\text{ReLU} }$.

\begin{equation}
    \hat{y} = \sigma_{\text{ReLU} } (\mathcal{W}\hat{x} + \mathcal{B})
    \label{eqn:nn-matrix}
\end{equation}
The weights, $\mathcal{W}_{ij}$ and biases, $\mathcal{B}_{i}$ are initialised uniformly following Glorot et. al. \cite{pmlr-v9-glorot10a}. The NN training phase is set to terminate early after seven consecutive epochs with no observable improvement in the $R^2$ loss value, $\epsilon$ as shown in Fig.\ref{fig:loss-vs-epoch}.

Additionally, 20\% of the data is set aside for validation tests; evaluating loss at the end of each epoch. They are selected from the last 20\% of the data before shuffling the rest. The accuracy over epochs of the NN training is evaluated with the coefficient of determination, $R^2$. It is the proportion of the variance in the dependent variable that is predictable from the independent variables.
\begin{equation}
    R^2 = 1 - \frac{\Sigma_i (y_i - f_i )^2 }{\Sigma_i (y_i - \bar{y} )^2 }
\end{equation}
and $\epsilon = 1-R^2$.

In the best case, the trained values, $f_i$ exactly match the true values $y_i$, which results in $\Sigma_i (y_i - f_i )^2 = 0$ and $R^2 = 1$. A baseline model, which always predicts $\bar{y}$, will have $R^2 = 0$. and negative values  when worse predictions are obtained.

\section{Closure test and prediction}
\begin{figure}[ht]
    \centering
    \subfloat[Protons]{
        \includegraphics[width=.9\columnwidth]{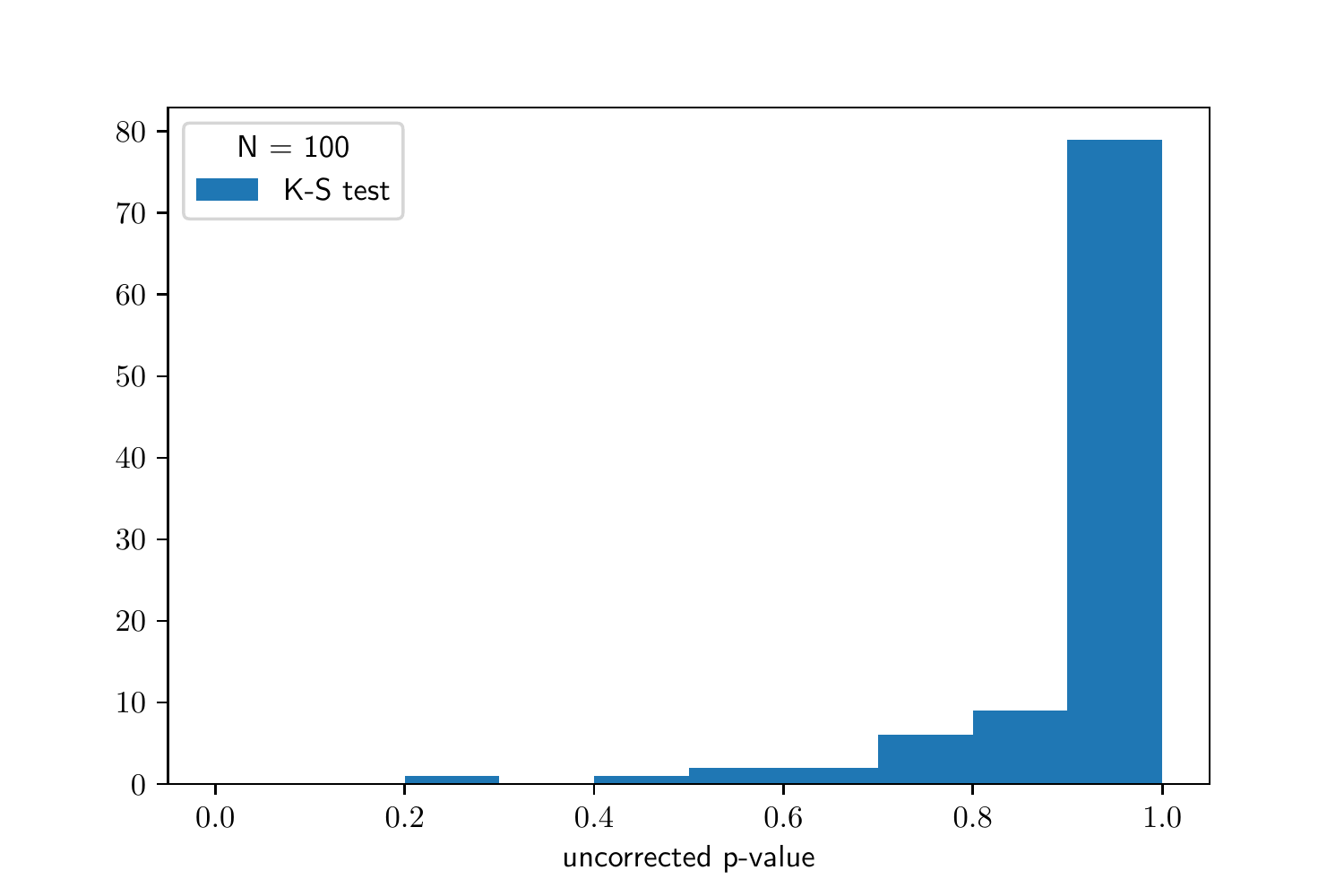} }
        
    \subfloat[Alpha particles]{
        \includegraphics[width=.9\columnwidth]{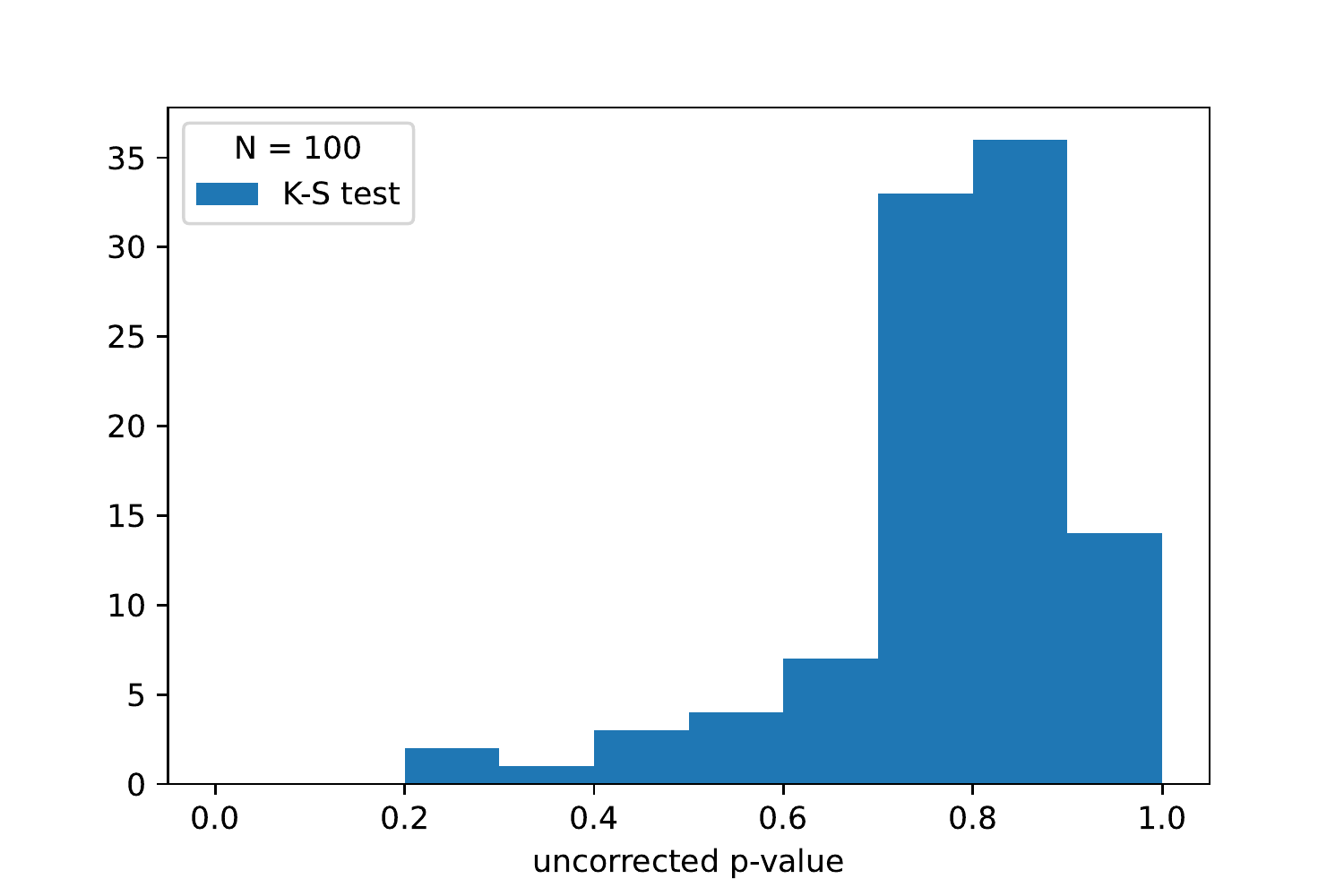} }
    \caption{The K-S p-values for bias estimation from NN unfolding. The results suggests no difference between the truth and unfolded distributions except due to statistical noise.}
    \label{fig:closure_tests}
\end{figure}

A standard way to estimate bias in a fitting algorithm is to conduct a closure test. Here, 100 independently generated data sets are used and the Kolmogorov-Smirnov (K-S) test p-value is obtained for each unfolded distribution. One expects to obtain unbiased results when unfolding observations drawn from the modified truth distribution. This is shown by the p-values obtained from each test data set as shown in Fig.~\ref{fig:closure_tests}. Only the energies, $E >$\SI{2.7}{MeV} for protons and $>$\SI{13.1}{MeV} for alpha particles are considered. This is because the detectors are no longer sensitive to the respective charged particles below those energies. Considering a standard $\alpha = 0.05$ and using the Holm-Bonferroni p-value correction, the unfolded data are statistically similar to the truth.

\begin{figure}[ht]
    \centering
    \subfloat[Protons]{
        \includegraphics[width=.9\columnwidth]{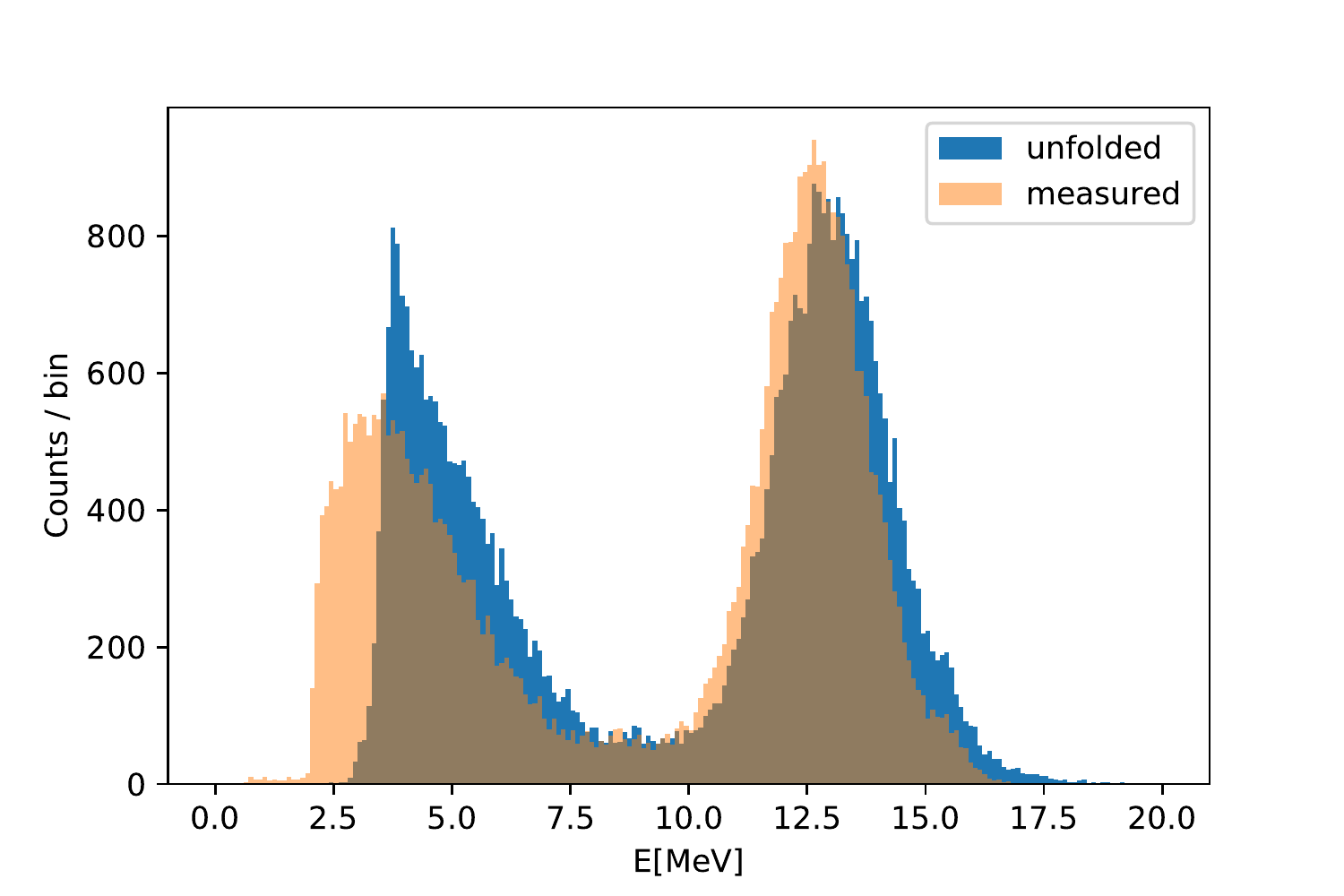} }
    
    \subfloat[Alpha particles]{
        \includegraphics[width=.9\columnwidth]{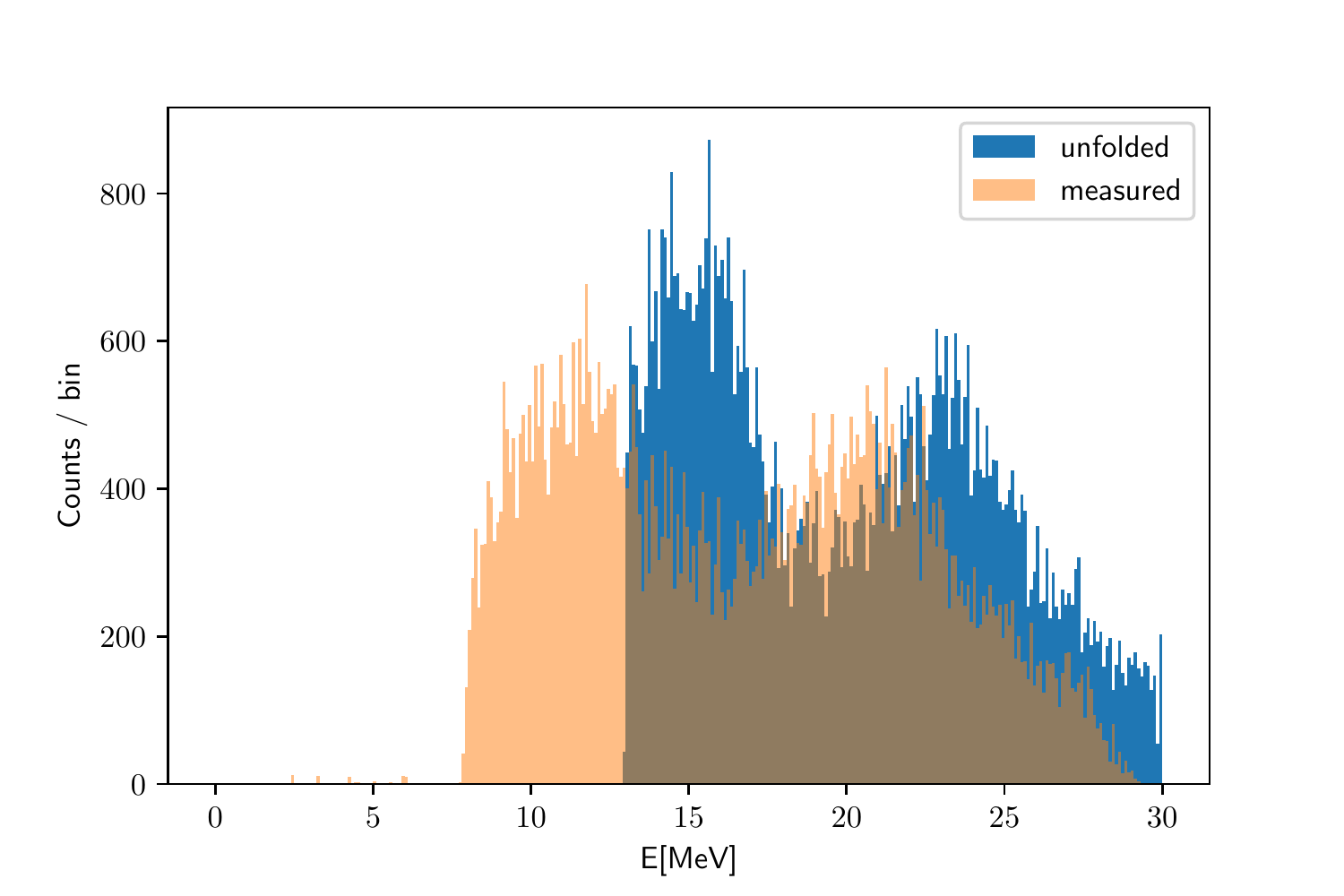} }
    \caption{Comparison between the measured and unfolded distribution.}
    \label{fig:gaussian-distribution}
\end{figure}

The performance of the NN is also tested with independent Monte Carlo simulations of the same setup geometry used during training but with a mixed Gaussian initial charged particle energy distribution. The trained model is used to predict the true distribution from the measured energy distribution as shown in Fig.\ref{fig:gaussian-distribution}. For protons, the two Gaussian distributions have means of \SI{4}{MeV} and \SI{13}{MeV} and sigmas of \SI{1.5}{MeV} and \SI{1}{MeV}. A simple background is modeled with a constant rate of 0.1 counts / bin. For alpha particles, the two Gaussian distributions are with means of \SI{14}{MeV} and \SI{23}{MeV} and sigmas of \SI{2.5}{MeV} and \SI{2}{MeV} and a background of 0.5 counts / bin.

\begin{figure}[ht]
    \centering
    \subfloat[Protons]{
        \includegraphics[width=0.9\columnwidth]{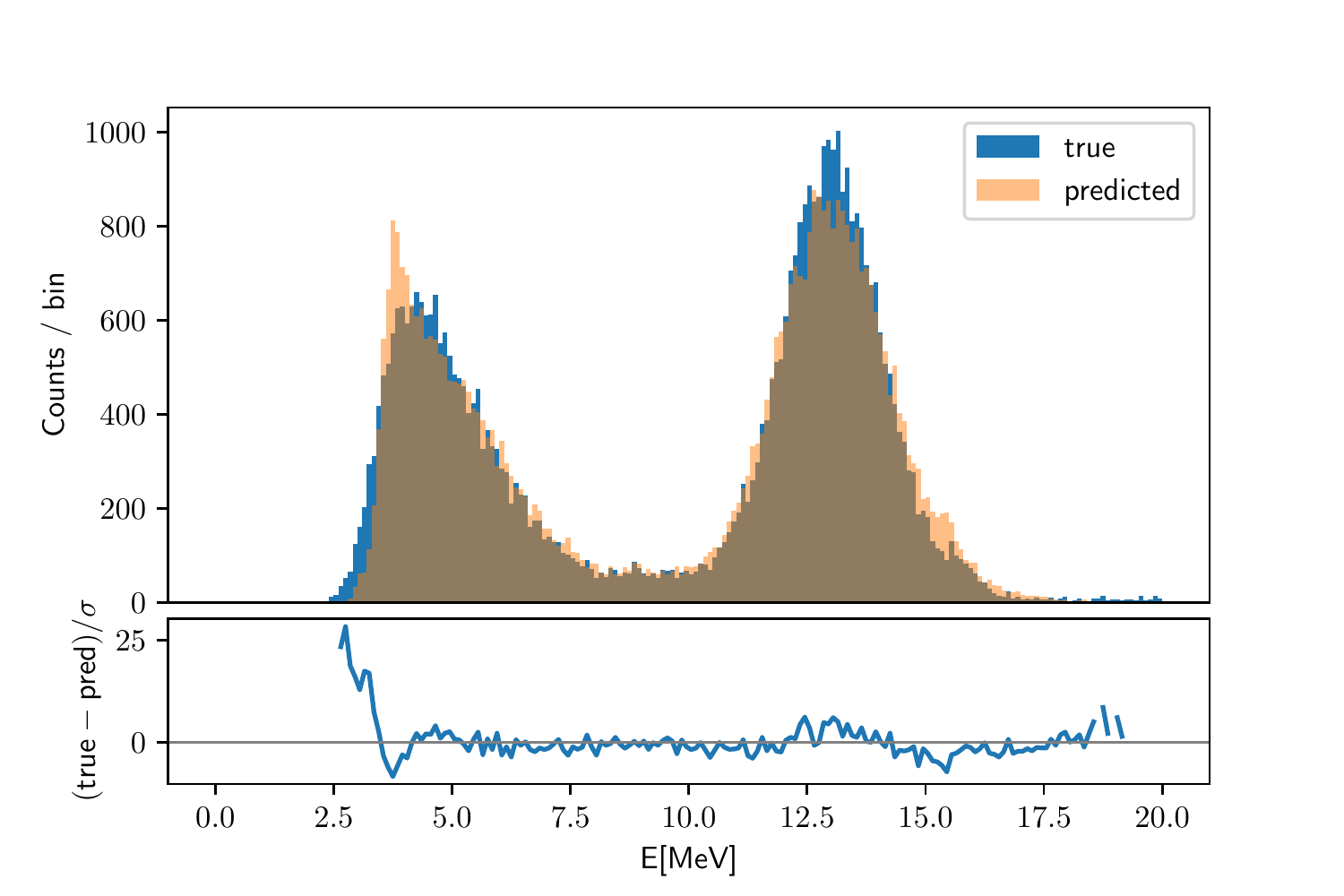} }
    
    \subfloat[Alpha particles]{
        \includegraphics[width=0.9\columnwidth]{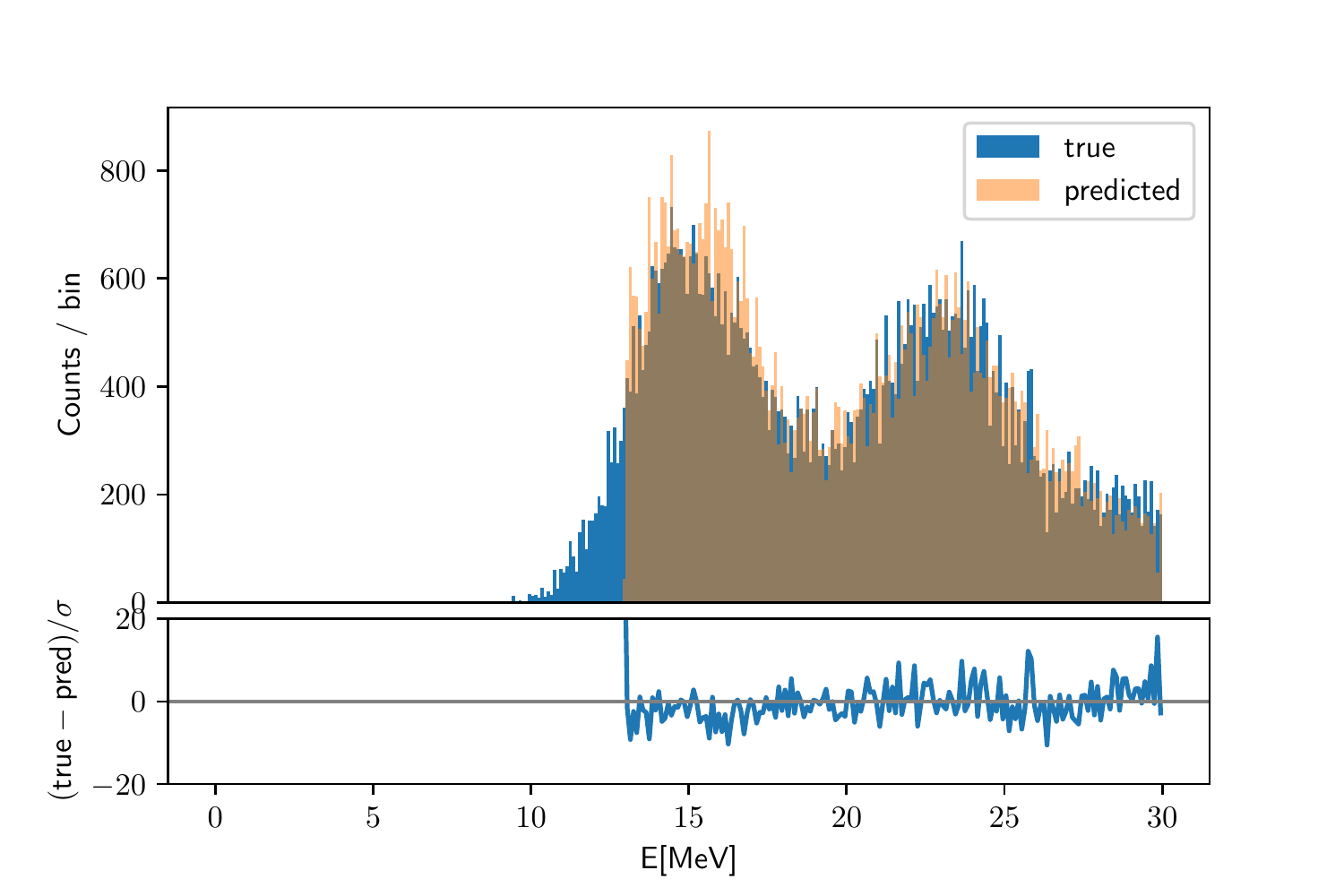} }
    \caption{The existence of a sharp dropoff at \SI{2.7}{MeV} for protons and \SI{13.1}{MeV} is because these charged particles were not able to pass through the first dE layer of the detector.}
    \label{fig:gaussian-comparison}
\end{figure}

The unfolding of experimental data is often complicated by a number of factors unrelated to the methodology itself, such as possible dataset inconsistencies or inadequacies of the theoretical description adopted and training statistics. For example, in this particular simulation setup, we are limited by the resolution, efficiency and energy range of the dE/dx detector. Therefore, it is less efficient to assess the benefits of a specific fitting methodology by applying it to the actual data, while it is much more robust to test it instead in an analysis of pseudo-data generated from a known prior. Therefore we test the NN with a known mixed Gaussian distribution as shown in Fig.\ref{fig:gaussian-comparison}. We can clearly see for alphas that the energy cut-off at around \SI{8}{MeV} is due to the fact that the dE/dx detector is sensitive only from this energy which may be unfolded to \SI{13}{MeV}.

\section{Hyperparameter tuning}\label{sec:hyperparameter_tuning}
The performance of different NN geometries are compared to understand the choice of hyperparameters that predicts the truth energy distribution. These hyperparameters include the choice of training epoch, number of layers, nodes and test-validation split ratio. The evaluation is done by repeating 3 times a 3-fold stratified cross validation randomized differently for each repetition. The performance will be evaluated using the previously mentioned $R^2$ for the training and test data. This cross-validation grid-search over a parameter grid is done with GridSearchCV\cite{scikit-learn} as it provides a convenient framework to keep track of the results of neural networks with different hyperparameters. 
\begin{figure}[ht]
    \centering
        \subfloat[Protons]{
            \includegraphics[width=.9\columnwidth]{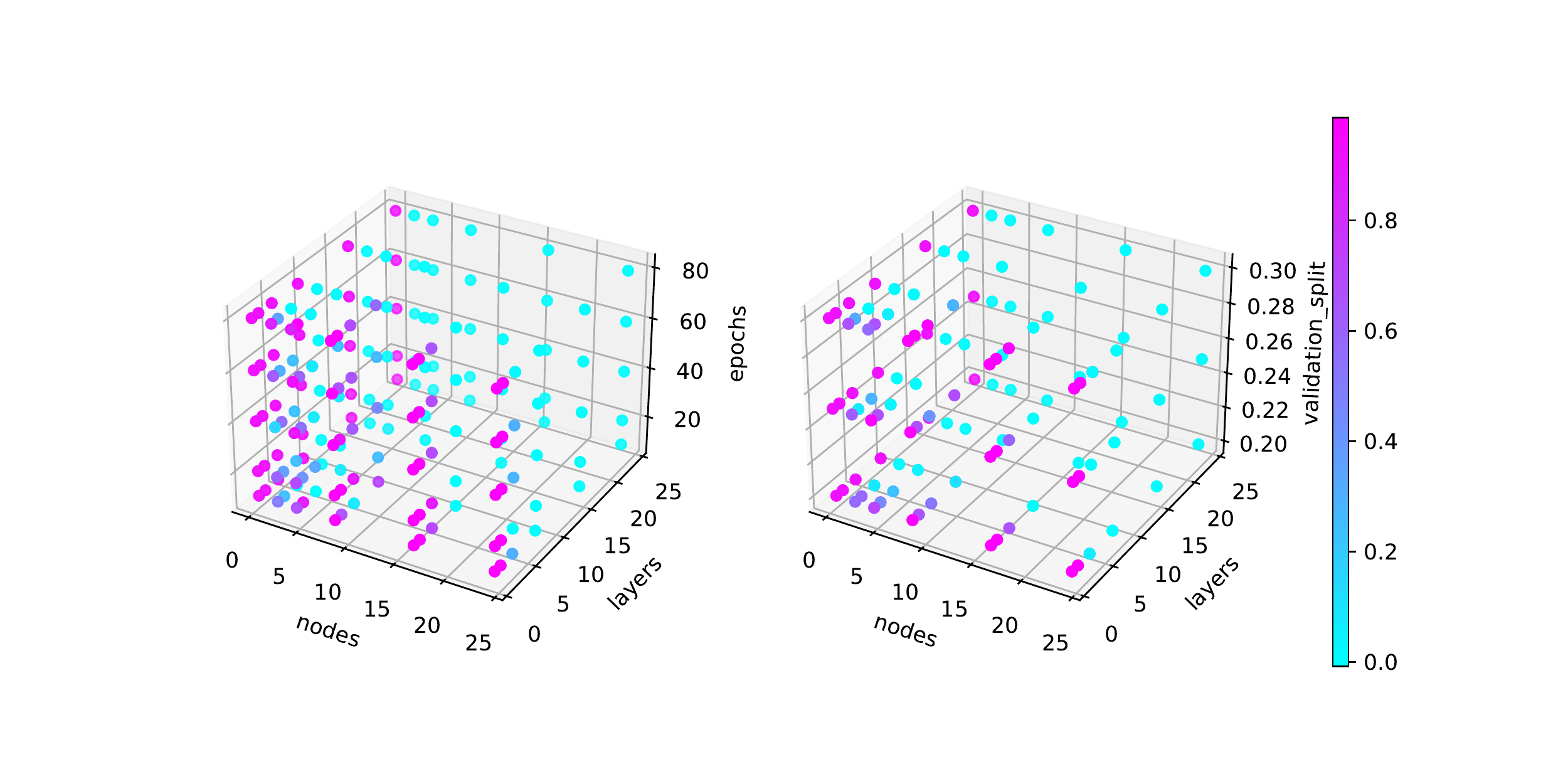} }
        
        \subfloat[Alpha particles]{
            \includegraphics[width=.9\columnwidth]{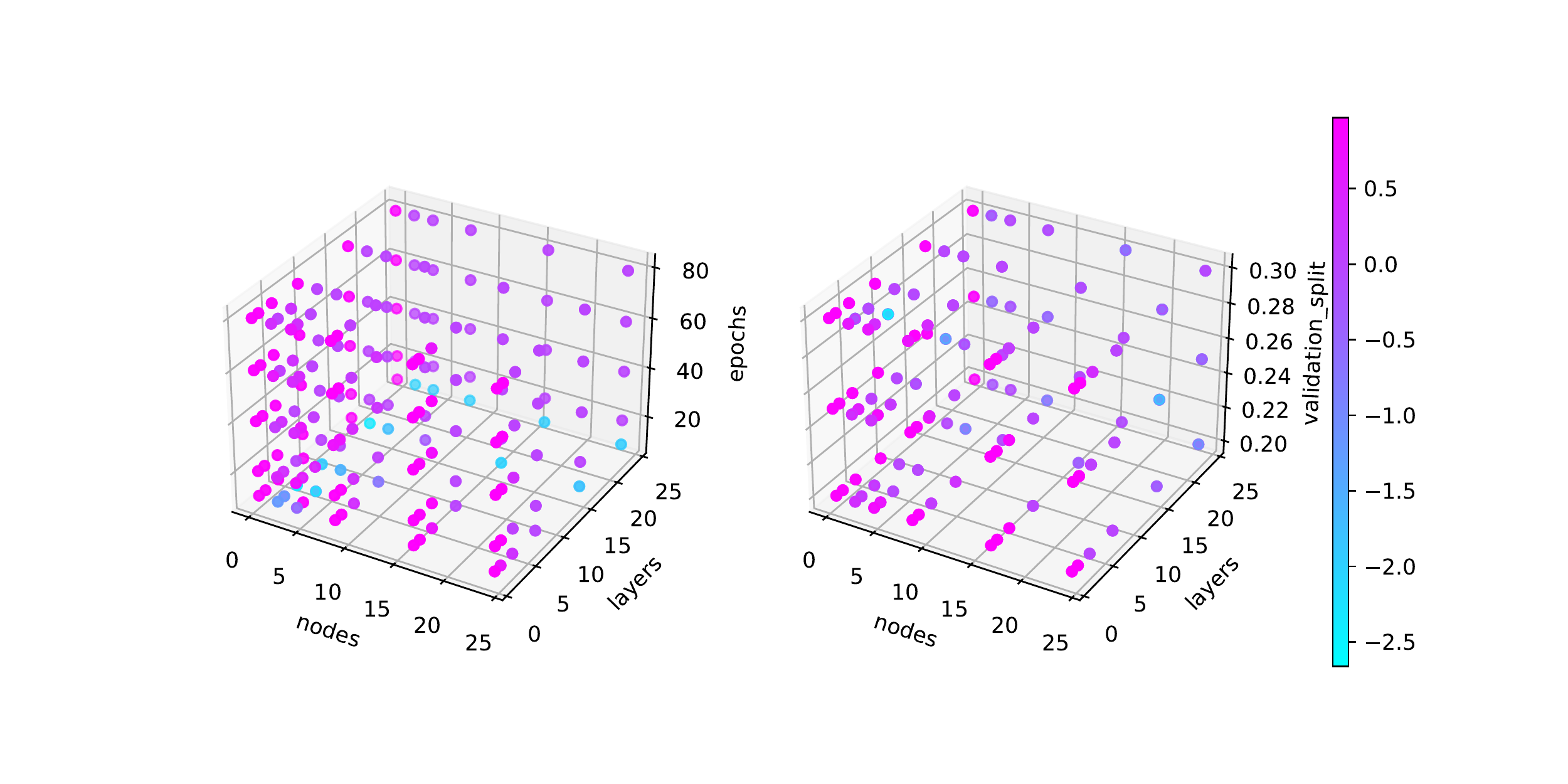} }
    \caption{$R^2$ scores for protons and alphas from several combinations of NN hyperparameters.}
    \label{fig:cv-scores}
\end{figure}
This automation enabled the possibility to test thousands of possible NN configurations and rank their results based on the $R^2$. A sample of this is shown in Fig.\ref{fig:cv-scores}. We then use the typical null hypothesis testing to determine if the top few different geometries are statistically different using Holm-Bonferroni corrected p-values. The corrected p-values for $R^2$ suggest that the top ranked NN geometries are statistically the same. The NN with a simpler geometry is chosen if the others are statistically similar.

\section{Comparison with other methods}
\begin{figure}[ht]
    \centering
    \subfloat[Protons] {
        \includegraphics[width=.9\columnwidth]{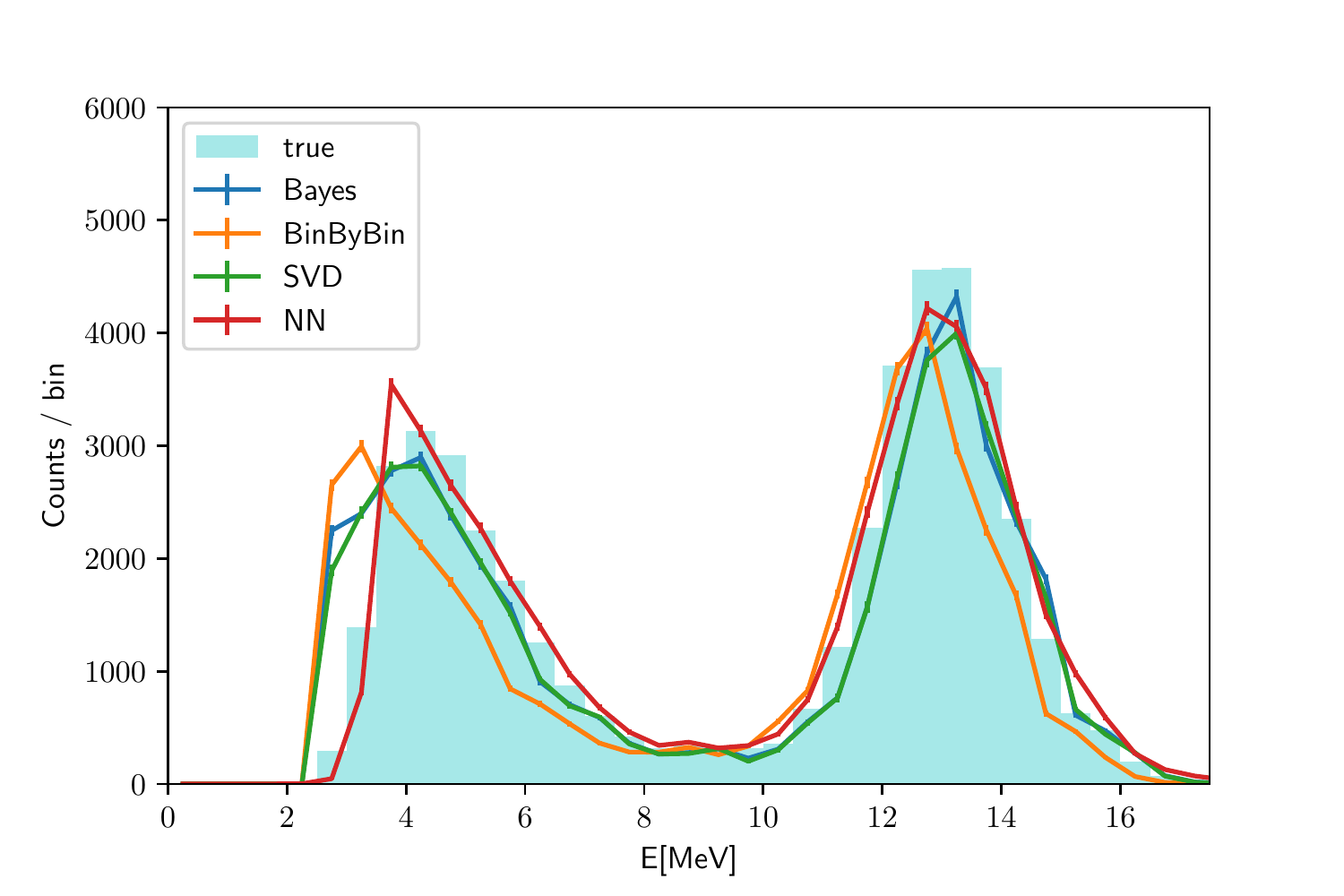} }
        
    \subfloat[Alpha particles] {
        \includegraphics[width=.9\columnwidth]{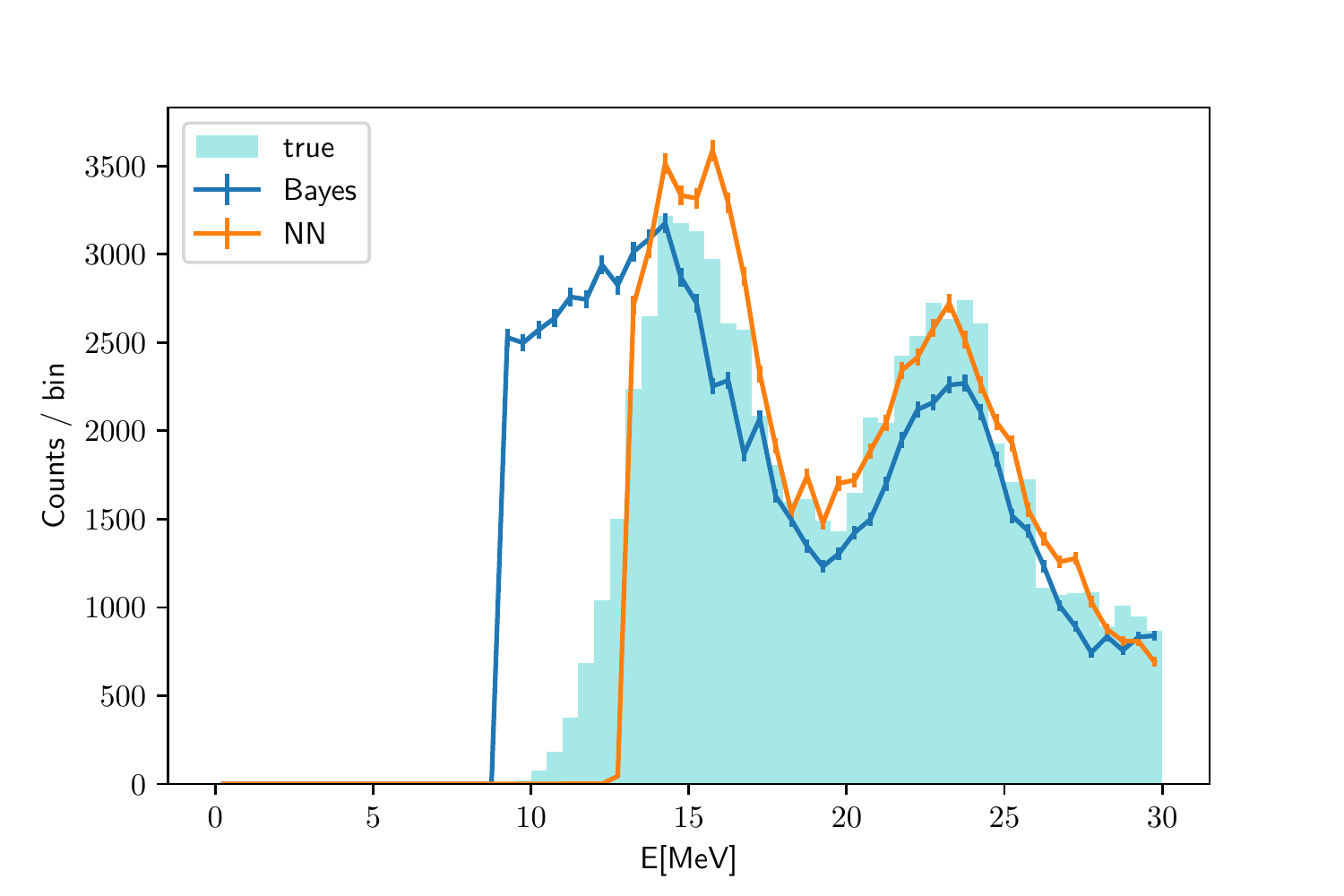} }
    \caption{Other more conventional techniques show larger discrepancies. For protons, there are large fluctuations for energies beyond \SI{16}{MeV} for SVD and d'Agostini Bayesian methods that are due to unfolding without simulated detector data so they are not shown. For alpha particles, the BinByBin and SVD methods produced wildly oscillating unfolded distributions and therefore are not shown.}
    \label{fig:proton-other-comparison}
\end{figure}

\begin{figure}[ht]
    \centering
    \subfloat[Protons] {
        \includegraphics[width=.9\columnwidth]{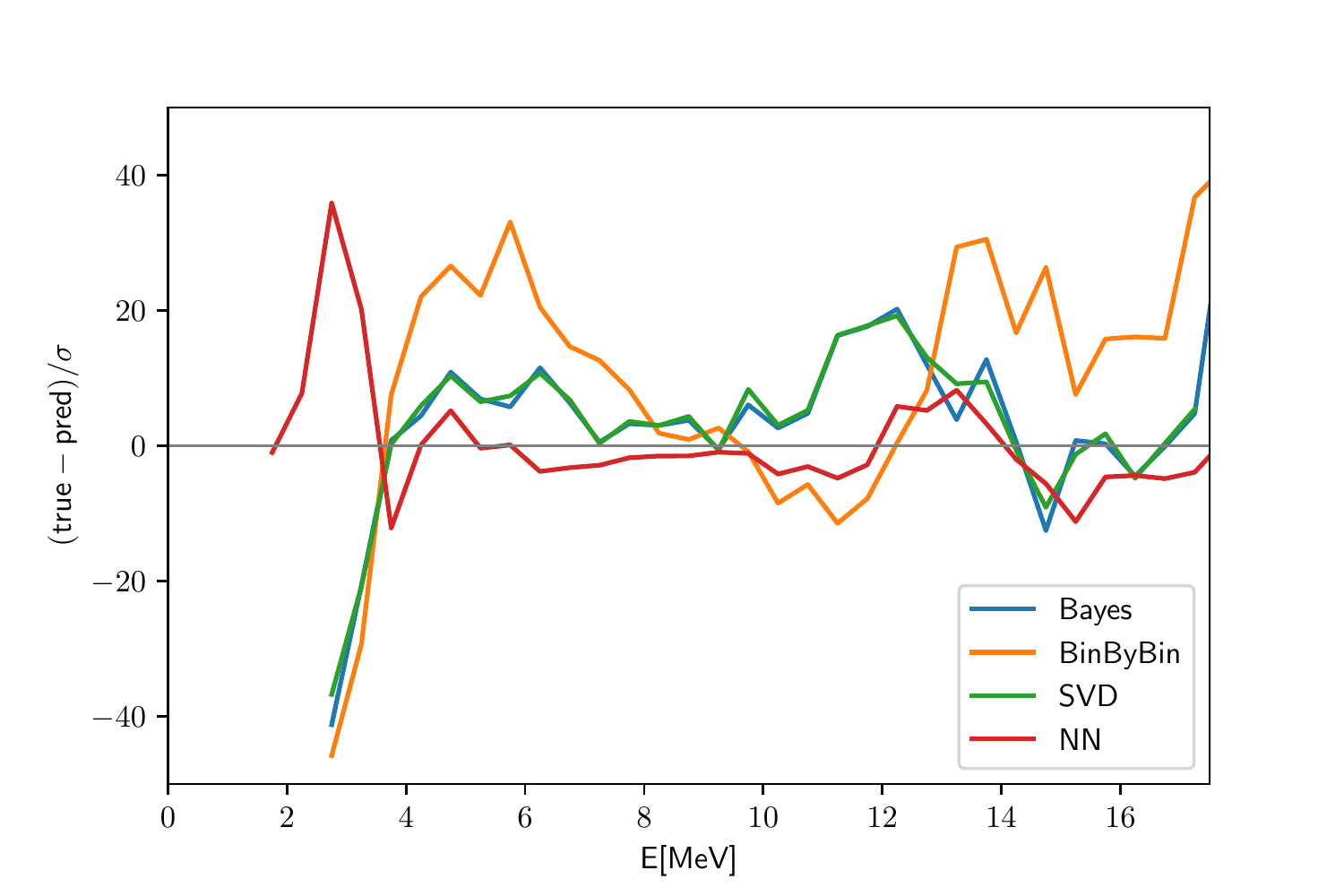} }
        
    \subfloat[Alpha particles] {
        \includegraphics[width=.9\columnwidth]{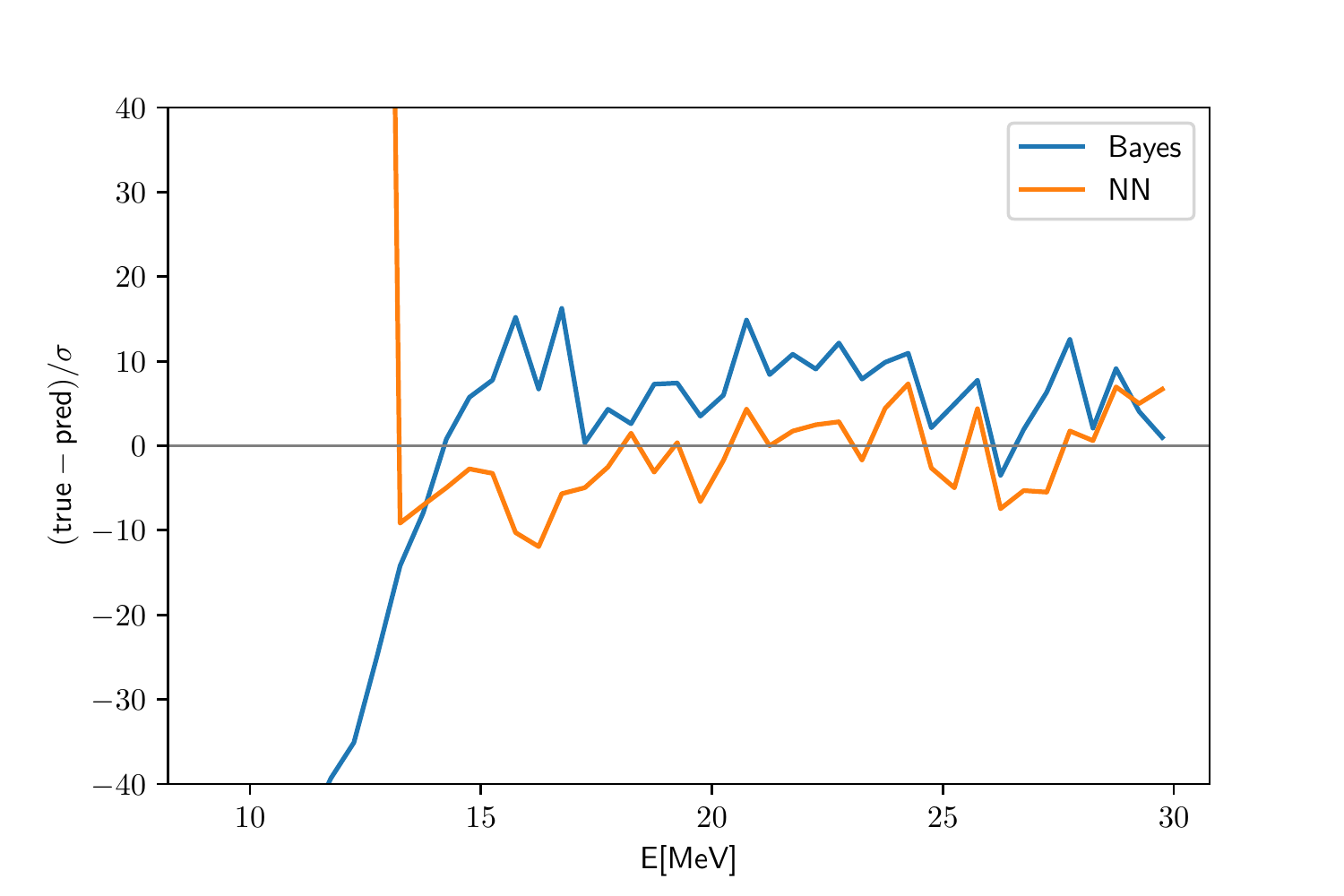} }
    \caption{Differences between the proton truth and unfolded distributions from various algorithms. The best result was possible with the use of NN. It has the least deviation from truth above \SI{3}{MeV}.}
    \label{fig:proton-comparison-diff}
\end{figure}

Currently, there are some unfolding algorithms in use like iterative d'Agostini Bayesian unfolding\cite{dagostini2010improved}, and SVD\cite{Hoecker_1996}. In Fig.\ref{fig:proton-other-comparison} and Fig.\ref{fig:proton-comparison-diff} the NN algorithm out-performs these methods. The data is prepared the same way except for additional step of constructing of a 50 x 50 response matrix that maps true to detector measured energies. A 60 x 60 response matrix was used for unfolding alphas as the alphas lose more energy over a wider energy range. The switch to \SI{0.5}{MeV} bin size is such that random fluctuations would not be an issue when performing this comparison. It is to be noted that experimentally there was an estimated 3.2\% and 4.1\% error for protons and alphas respectively due to energy miscalibration and geometrical uncertainties. The uncertainty from the NN is purely statistical that depends on the kernel initialization. Other techniques do not perform as well as the NN  as it could also recover the truth distribution for a wider energy range.

Other unfolding methods also exist, such as a least square fit with Tikhonov regularisation\cite{Schmitt2012}, Monte Carlo events re-weighting\cite{LINDEMANN1995516} and particle swarm optimization\cite{SHAHABINEJAD20179} but are not included in this study.

\section{Summary}
It is shown that the neural network can efficiently produce unfolded distributions that approximate the truth distribution very well. The simple network geometry is tested with unfolding protons and alphas generated from MC and both passed closure tests and predicting truth data. Finally the comparison with other more established methods such as the d'Agostini Bayesian iterative and SVD methods show that the NN works better and does not require further regularisation.

\section{Acknowledgements}
We would like to thank the AlCap collaboration for support.

\bibliographystyle{elsarticle-num}
\bibliography{references}

\end{document}